# On hydrogen bond correlations at high pressures

S.K. Sikka

Office of the Principal Scientific Adviser to the Government of India, 324, Vigyan Bhawan Annexe, Maulana Azad Road, New Delhi 110011, India, Email: sksikka@nic.in

Abstract

In situ high pressure neutron diffraction measured lengths of O . H and H …O pairs in hydrogen bonds in substances are shown to follow the correlation between them established from 0.1 MPa data on different chemical compounds. In particular, the conclusion by Nelmes et al that their high pressure data on ice VIII differ from it is not supported. For compounds in which the O . H stretching frequencies red shift under pressure, it is shown that wherever structural data is available, they follow the stretching frequency versus H . O (or O …O) distance correlation. For compounds displaying blue shifts with pressure an analogy appears to exist with 'improper' hydrogen bonds.

It is well known that at normal pressure, neutron diffraction and spectroscopic data on different chemical substances for X-H ---A hydrogen bonds show two major correlations with decreasing H---A or X---A distances: (i) a lengthening of the covalent X-H bond and (ii) a decrease in the frequency ($\nu_{X-H}$) of the X-H stretching vibration [1]. Application of high pressure is expected to reduce the interatomic distances in a given substance. Since the compressibility of a covalent bond is small, a decrease of X-H---A distance or H---A, from correlation (i), thus should lead to an increase in the X-H bond length. This indeed has been observed in some neutron diffraction studies done under pressure. In particular, Nelmes et al [2] measured O-H distances in ice VIII up to 10 GPa and found that its rate of increase with pressure was 4 (4) * $10^{-4}$ Å /GPa. This result was confirmed by ab initio Hartee Fock calculations done by Besson et al [3] and Ojamae et el [4]. The former gave the value 4 (4) * $10^{-4}$ Å /GPa while Ojamae at el put this value at 1.4 * $10^{-4}$ Å /GPa. (Subsequent measurements by Nelmes et al [2] on ice to 25 GPa placed the value of the rate at 6.3 (1.2) * $10^{-4}$ Å /GPa). For NaOD-V, Loveday et al [5] obtained $\delta r_{(O-H)}/\delta P$ to be 2 (4.9)* $10^{-4}$ Å /GPa. No significant variation in O-H distances with pressure has also been reported in hdrogarnet [6] and chondrodite [7].

However, the above-observed rates of increase were much smaller than predicted values (20 to 30 * $10^{-4}$ Å /GPa.) from some earlier empirical calculations and from fits to observed changes of O-H vibrational frequencies under pressure. Since, the authors of those studies had assumed that the O-H distance varies with O---O distance with pressure similar to that, known at time, for the chemical compounds at normal pressure, Nelmes et al [2] concluded that the small observed rates represented deviation between the high pressure behaviour of O-H distances from that derived from chemical substances, implying that the form of the hydrogen bond potential would not be constant with pressure. This is widely accepted in literature.

In 1997, Sikka [8] examined this question of the dependence of the X-H distance of hydrogen bond on pressure through a modification of a very successful Lippincott and Schroeder potential function for an isolated hydrogen bond (Chidambaram and Sikka [9] ). For ice, this modification consisted of a introduction of an additional term to take care of increased repulsion felt by the hydrogen atom from its neighbouring hydrogen atom in the cooperative hydrogen bond in ice (- O-H---O---H-O - ). However, the difference in r values with and without this repulsive term was only 0.0005 Å muck less than the claimed precision ( 0.003 Å ) of the neutron experiments under pressure. Applications of density functional theories (DFT) since then have provided very precise values of the hydrogen bond parameters. A linear fit to his ab initio results by Fortes [10] yielded a value for $\delta r_{(O-H)}/\delta P$ to be 9.4 (0.3)* $10^{-4}$ Å /GPa. He states that given the scatter in experimental data, the agreement between the experiments and theory is satisfactory.

Then where is the discrepancy? Are the estimates of $\delta r_{(O-H)}/\delta P$ used to compare experimental values suspect? A very simple analysis can be done to test these. This is by directly plotting the high pressure experimental O-H against H---O values on the graph, assembled from data on the chemical substances at 0.1 MPa. This is shown in Fig.1. The plot in the figure contains data from 48 organic compounds and 20 hydrates. The data have been taken from Cambridge Data base of 2004 [11]. Only structural analysis done at low temperatures ( < 125 K) and with R values < 0.06 were considered. Also O-H distances only above 0.94 Å were accepted. The comparison shows that the high pressure data are consistent with chemical data in the region of comparison and points out that the experimental data of Nelmes et al [2] on ice are essentially correct. The data for NaOH-V[5], chonrodrite[7], clinochlore[12] and kalcinite[13] are also in agreement with the chemical data. The hydrogen bonds here are very weak and not much variation in O-H distances is expected according to the chemical plot. Till date, no neutron diffraction measurements have been possible in the pressure regime where a faster variation of O-H should occur against O---O lengths.

The changes in vibration frequencies with pressure are relatively easier to monitor and measureable to higher pressures than structural parameters in neutron diffraction experiments. Three types of behaviour of O-H stretching frequency with pressure have been observed: (i) softening (red shift) with pressure, (ii) almost no variation with pressure and (iii) positive shift (blue shift) with pressure. In case (i), where structural data is available, this downward shift of the O-H stretching modes is consistent with that observed for chemical substances at ambient pressure. For example, figures 2 and 3 display this for $M(OH)_2$ oxides [14,15] and ice [17]. For case (ii), some examples are NaOH-V [18], hydrogarnet [6] and kalcinite [13]. Here we note from their structural data that either the H---O distances in them are larger than 2 Å or are constant with pressure. For the first case, Fig.2 shows that $\upsilon_{X-H}$ does not vary much up to this distance. Hofmeister et al [19] have shown that changes in stretching frequencies with pressure not only depend on X-H---A distances but also on the X-H---A angles also. However, the effect of the angle can be better represented by that of the H---A distance. For example, in kalcinite [13], both the O-H---O distances and the O-H---O angles decrease with pressure but the H---O distances are invariant. So are the O-H stretching frequencies.

For case (iii), Table 1 gives some examples. The blue shifting of stretching frequencies occurs very near the value ~ 3600 $cm^{-1}$. Here, in some cases, one may not associate these with hydrogen bonds as the H---O distances are larger than 2.31 Å, the cutoff distance for hydrogen bonds set by Klein [24], based from electron density topology considerations. Notwithstanding this, the O-H potential in such cases will be shallow (with dissociation energies of ≤ 1 kilocal /mol) and an entire range of bonding configurations is possible as the crystal packing can now easily bend, elongate or compress the bond [1].

Can the O-H---O bonds in them be regarded as belonging to a class of hydrogen bonds called "improper" ones [25]? At normal pressure, the stretching frequencies of the latter are blue shifted with respect to isolated or free ion frequencies. Although blue shifts have been reported mainly for C-H bonds, for O-H---Y bonds, these have also been known for O-H ions complexes with metal ions (for a discussion of this see Hermansson [26]). Recent theoretical calculations by Alabugin et al [27] show that O-H---Y with Y as Ne and fluorine in $CF_4$ would have these 'improper' hydrogen bonds. These calculations also show that a blue shift means an X-H contraction. This is ~ .0004 Å to 0.001 Å (see Tables 9 and 10 of Alabugin et al [27]). Such small changes in X-H distances may cause frequency changes ~ 10-40 $cm^{-1}$ [28].

Theories for the formation of improper hydrogen bonds are still under debate [27,29]. Steric effects in a structure have been shown to be one of the causes of blue-shifting hydrogen bonds [30]. It may then be not surprising that in materials under pressure, because of increasing crowding of atoms, blue-shifting hydrogen bonds should be found. Following may be noted about the substances in Table 1.

Interestingly, two of the compounds in Table 1 are F-bearing, i.e., there is a partial substitution of O-H by F atoms. Further, while the F-bearing topaz hydroxyl shows a blue shift of an O-H stretching frequency, pure topaz O-H shows the usual red shifts [31]. Comparison of structural data may clarify this.

In chondrodite-OH/D(F)[7] and clinochlore [23], O-H distances have been measured under pressure. In the former, the O-D distance is found to be almost constant up to 5.27 GPa and decreases from 0.96 Å to 0.93 Å at the next reported pressure value of 7.04 GPa However, this change is too large compared to the expected values mentioned above. Thus, this may be due to an artifact in the experiment. We may note that the current accuracy of determining X-H distances by neutron powder method under pressure is not less than 0.02 Å. In clinochlore, the O-D distances are almost constant up the pressure of measurement of 4.7 GPa.

The O-H bond geometry of clinohumite and chrondrodite is very similar [32] (see Fig.4). Here, the M-H distances are quite short (in chrondrodite these are: 2.36-2.47 Å, compared to the usual value of about 2.7 Å) and thus there is additional repulsion between $H^{+\delta}$ and $M^+$ atoms. This may produce a slight lengthing of the O-H bond, leading to the blue shifts. Again in clinochlore, there are short Mg-D1 (2.57 Å) and Al-D2 (2.58 Å) distances.

Now we discuss the case for dense, hydrous Mg silicates phases, A, B and superhydrous B (Shy B). In phase A, Hofmeister et al [19] observed positive shifts with pressure for the 3518 $cm^{-1}$ frequency. However, in Raman studies [33], negative shifts have been observed for the same frequency (Here, we do not discount the possibility of different pressure responses in Raman and IR studies). In phase B, the frequency at 3410 cm$^{-1}$ first shows negative shifts and then positive shifts from about 5 GPa. In ShyB phase, positive shifts for the frequency 3411 cm$^{-1}$ have been observed with pressure. It may be noted that these compounds show short repulsive H-H contacts, which appear to be ordered unlike in chondrodite, clinohumite and hydroxyl topaz [34] where these are disordered. (the contacts, here, are no longer repulsive as in the actual structure the hydrogen atoms are never in close proximity). In 1997, Sikka [8] calculated a contraction of the O-H distance for ordered H-H contacts. This was of the same order of magnitude as mentioned above for improper hydrogen bonds.

From the above discussion it is clear that more work needs to be done to quantitatively understand these improper hydrogen bonds under pressure.

Thanks are due to Dr.R. Chitra of Solid State Physics Division of Bhabha Atomic Center ,Mumbai for providing data from Cambridge Data Base.

Table 1: Some minerals where the O-H stretch mode is blue shifted under pressure

| | $\upsilon_0(cm^{-1})$ (p=0.1MPa) | sign of $\delta\upsilon/\delta p$ | H$\cdots$O(Å) | Reference |
|---|---|---|---|---|
| topaz-OH(F) | 3650 | + | 2.28 | [20] |
| clinohumite-OH | 3607 | + | 2.54 | [21] |
| | 3561 | + | | |
| | 3525 | + | | |
| choridrodite-OH(F) | 3688 | + | 2.57 | [7, 22] |
| | 3566 | + | | |
| | 3558 | + | | |
| | 3383 | - | 1.92 | |
| clinchlore* | 3679 | + | no-H bond | [12, 23] |
| | 3647 | 0 | 2.16 | |
| | 3605 | + | 2.16 | |
| | 3477 | 0 | 1.88 | |

\* Raman data

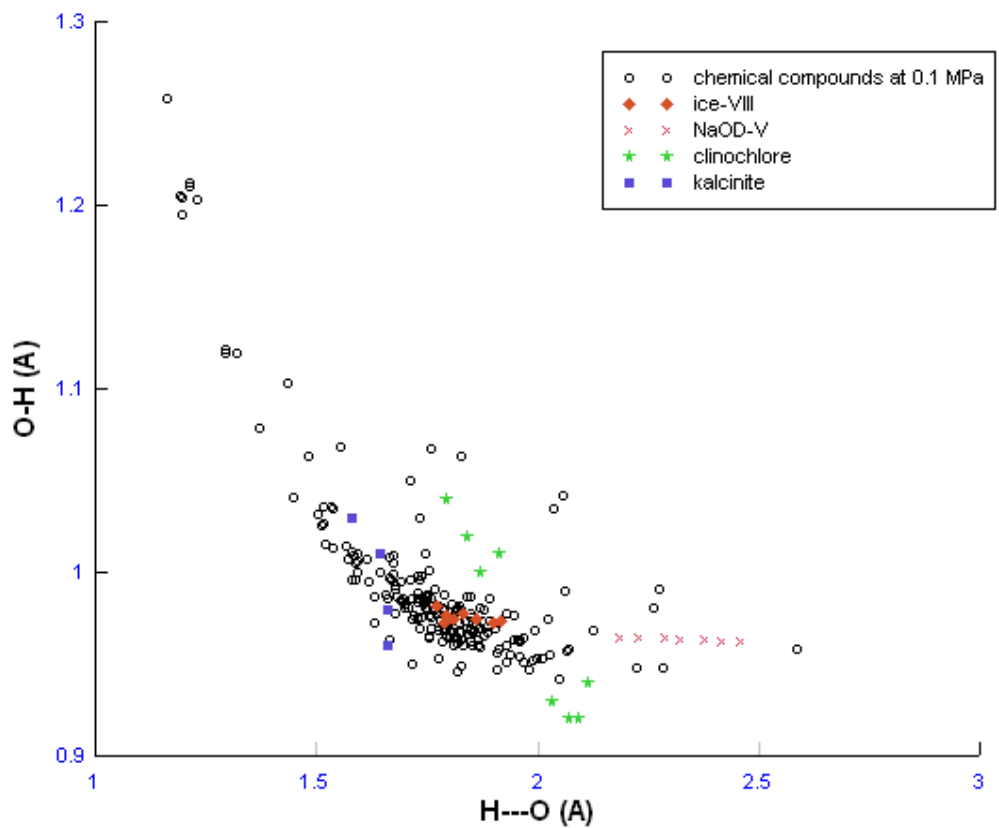

Fig.1  Correlation of the O-H and H---O distances in O-H---O hydrogen bonds.

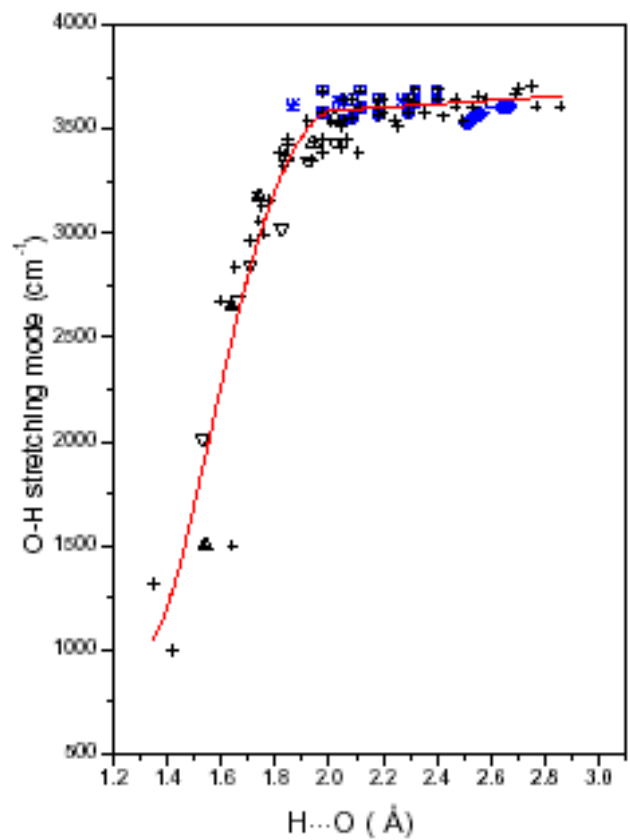

Fig.2 Frequencies of the O—H stretching mode versus H⋯O distance, Blue colour symbols represent high pressure data of M(OH)$_2$ oxides. The curve is fit to the eye of 0.1 MPa data assembled by Libowitzky [16] on different minerals.

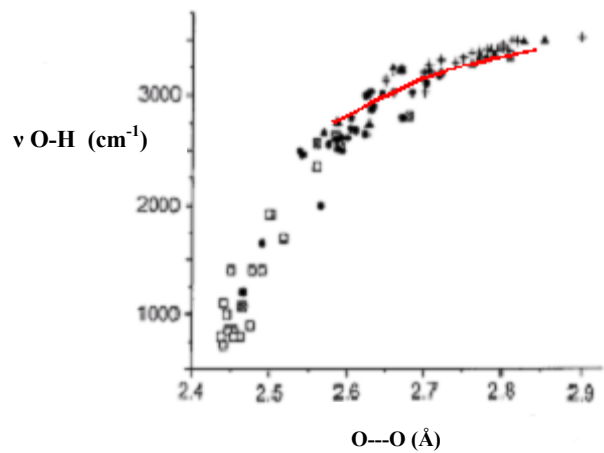

Fig.3 Correlation of IR stretching frequencies against O----O distances in O-H---O hydrogen bonds taken from [1]. The red curve represents the experimental data for ice for pressures before interference from Fermi resonance.

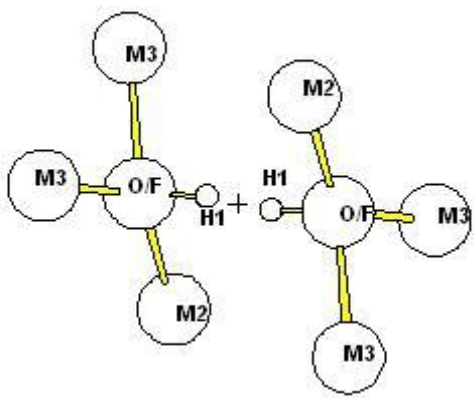

Fig.4 The O-H bond geometry of chondrodite-OH(F) and clinohumite-OH(F). M atoms are mostly Mg. The occupancies of the two centrosymmerically related H sites is half or less depending upon the F content.